# General theory for super-sensitive dual-wavelength phase metrology: error-free unwrapping and signal-to-noise ratio

**Manuel Servin,**[*] **Moises Padilla, and Guillermo Garnica**

*Centro de Investigaciones en Optica A. C., 37000 Leon Guanajuato, Mexico.*
*\*mservin@cio.mx*

**Abstract:** From 1971 to 2012 dual-wavelength optical-metrology used only the demodulated low-sensitivity phase-difference of two close-sensitive fringes. Dual-wavelength phase-metrology that additionally uses the phase-sum was first reported by Di et al. in 2013 [28]; this was an important step to increase the phase-accuracy in optical metrology. This and its derived papers however do not offer mathematical analysis for signal-to-noise ratio (SNR) for the phase-difference and phase-sum. Neither provide the mathematical analysis for unwrapping the phase-sum without errors. Here a general theory for super-sensitive two-wavelength phase-metrology is given. In particular mathematical analysis and formulas for SNR and error-free phase-unwrapping for two-wavelength metrology is provided. We start by phase-demodulating two close-sensitivity fringes by phase-shifting algorithms (PSAs). We then calculate their phase-difference and their phase-sum; the phase-difference is assumed non-wrapped. However the phase-sum is highly wrapped, super-sensitive and has much higher SNR. Spatial phase unwrapping for a highly discontinuous phase-sum is precluded. However as we show, it is possible to unwrap the noisy phase-sum from the noisier phase-difference without errors. We apply this super-sensitive phase-metrology theory to profilometry allowing us to obtain super-sensitive height measurements. To the best of our knowledge the mathematical analysis and formulas herein presented for the SNR and error-free unwrapping have not been reported before.

## 1. Introduction

Here we start by presenting the well known concept of synthetic lambda interferometry concept. Wyant used two close laser-wavelengths $\{\lambda_1, \lambda_2\}$ to test optical surfaces with a synthetic wavelength of $\Lambda_D = \lambda_1\lambda_2 / |\lambda_1 - \lambda_2|$ [3]. Given that $\lambda_1 \approx \lambda_2$, the wavelength $\Lambda_D$ is much longer than either $\{\lambda_1, \lambda_2\}$, *i.e.* $\Lambda_D >> \{\lambda_1, \lambda_2\}$. The wavelength $\Lambda_D$ is obtained by the difference $\varphi_D = \varphi_1 - \varphi_2$; being $\{\varphi_1 = (2\pi/\lambda_1)w, \varphi_2 = (2\pi/\lambda_2)w\}$, and $w(x,y)$ the wavefront under test. Double-wavelength phase-metrology was improved by Polhemus [4] and later on by Cheng [5,6] using digital phase-shifting algorithms (PSA). Afterwards Onodera et al. used Fourier phase-demodulation for profiling structures with long equivalent-lambda depth size [7]. Unfortunately, the demodulated phase was over-smoothed due to over filtered diffraction orders [7]. This in turn was followed by a large number of double-wavelength Fourier and PSA demodulation methods in such diverse applications as double-wave holographic microscopy [8], extended range optical metrology [9], two-step digital holography [10], multiwavelength extended-range contouring [11], and two-wavelength surface profiling [12]. Dual-wavelength optical metrology using $\varphi_D = \varphi_1 - \varphi_2$ has also been applied to holography, fringe-projection profilometry and contouring [13-20]. These methods have been called among other, two-frequency profilometry [13], profilometry without phase unwrapping [14], dual-frequency shape measurement [15], large-depth discontinuous objects profilometry [16], deflectometry of composite fringe phase retrieval [17], absolute-height fringe-projection profilometry [18], dual-wavelength two-steps phase-shifting demodulation [19], two fringe patterns absolute-phase recovery [20]. Even though these methods [3-20] were applied to different optical phase-metrology experiments, all of them share the same mathematical background just described. These authors referred their techniques as extended-range, absolute-phase measurement, phase-metrology without phase-unwrapping or direct shape measurement [3-20]. Finally we comment that the paper by Katherine Creath [27] (mentioned by one reviewer) also use just the phase-difference $\varphi_D$ in his paper. Creath do not mention the possibility of using the phase-sum $\varphi_2 + \varphi_1$ in his phase-metrology technique [27].

Therefore from 1971 [3] until 2012, dual-wavelength optical metrology was synonymous of demodulating two close-sensitive phases $\{\varphi_1, \varphi_2\}$ ($\varphi_1 \approx \varphi_2$) for obtaining the non-wrapped phase-difference $(\varphi_2 - \varphi_1) \in (-\pi, \pi)$. However in 2013 Di et al. [28] computed for the first time, the phase-sum $\varphi_S = \varphi_2 + \varphi_1$ to obtain higher sensitivity and higher SNR phase measurements. As 2017, both $\varphi_D = \varphi_2 - \varphi_1$ and $\varphi_S = \varphi_2 + \varphi_1$ for phase metrology has been published (in peer reviewed journals) by: Di et al. [28,29] (2013 and 2015) and Xiong et al. [31] (2017). And by Servin et al. in the arXiv.org [32], not knowing the existence of [28,29,31]. However searching in Google Scholar, the paper [28] has been cited just twice [29,31], so this paper is hardly known even today, four years later from its publication. This is unfortunate because we consider that [28] is an important leap forward in double-wavelength phase-metrology. That is because the phase-sum is far more sensitive than the phase-difference $\varphi_S = G \varphi_D$; being $G \gg 1.0$. And as we mathematically prove here, their SNR between $\{\varphi_S, \varphi_D\}$ is also much higher $\text{SNR}(\varphi_S) = G^2 \text{SNR}(\varphi_D)$.

On the other hand, Servin et al. [24,25], and afterwards Xiong et al. [31] showed that a higher sensitive phase may be unwrapped directly from a lower-sensitive non-wrapped one. Di et al. dealt only with continuous wavefronts $w(x,y)$ so they could unwrap $\varphi_S$ spatially [28,29]. Finally a reviewer called our attention to Zuo et al. [30], but this paper only unwraps $\varphi_D = \varphi_2 - \varphi_1$ (Eqs. (7)-(8)-(10) in [30]); they do not mention $\varphi_S = \varphi_2 + \varphi_1$.

Here we present a new results not presented in [28,29,31]. These are
- Mathematical analysis and a closed-form formula to obtain error-free unwrapping for the noisy phase-sum $\varphi_S + n_S$ directly from the noisy phase-difference $\varphi_D + n_D$.
- Mathematical analysis and closed-form formula for the SNR for noisy phase-sum $\varphi_S + n_S$ and phase-difference $\varphi_D + n_D$.

Finally we stress that $\varphi_S = \varphi_2 + \varphi_1$ has a higher sensitivity than either $\{\varphi_1, \varphi_2\}$ (super-sensitivity). We thank a reviewer for calling our attention to the recent work by Di et al. [29], and by using Google Scholar we found the two additional works which use $\varphi_S = \varphi_2 + \varphi_1$ [28,31].

The plan of this paper is as follows: section 2 reviews the state of the art in two-wavelength phase-metrology up to mid 2017. Section 3 we demodulate two close-sensitivity phases $\{\varphi_1, \varphi_2\}$, from which we compute $\varphi_D = \varphi_2 - \varphi_1$ and $\varphi_S = \varphi_1 + \varphi_2$. In section 4, we mathematically analyze the SNR for $\{\varphi_S, \varphi_D\}$. In section 5, we give a close-form formula for error-free unwrapping the noisy $\varphi_1 + \varphi_2$ from the noisy $\varphi_2 - \varphi_1$. In section 6 we use fringe-projection profilometry to exemplify our two-wavelength phase-metrology theory. In section 7 we show that $\varphi_1 + \varphi_2$ has about twice the sensitivity than a single high-carrier fringe-pattern. Finally in section 8 we draw some conclusions.

## 2. State of the art in two-wavelength optical metrology

As mentioned, during 1971-2012 dual-wavelength phase-metrology used only $\varphi_D = \varphi_2 - \varphi_1$ from two close-sensitive phases $\{\varphi_1, \varphi_2\}$; being $\varphi_D \in (-\pi, \pi)$ [3-27]. In 2013 Di et al. [28] first used the higher-sensitive $\varphi_S = \varphi_2 + \varphi_1$ for dual-wavelength phase-metrology; this was an important leap forward in phase-precision in phase-metrology [28].

Let us consider two close-wavelengths $\lambda_1 \approx \lambda_2$ (close-sensitivity) fringe patters as,

$$I_1(x,y,\Delta) = a(x,y) + b(x,y)\cos\left[\frac{2\pi}{\lambda_1}w(x,y) + \Delta\right];$$
$$I_2(x,y,\Delta) = a(x,y) + b(x,y)\cos\left[\frac{2\pi}{\lambda_2}w(x,y) + \Delta\right]. \quad (1)$$

Where $a(x,y)$ is the background; $b(x,y)$ the contrast; $w(x,y)$ the measuring wavefront, and $\Delta$ a piston phase. From Eq. (1) we may find two synthetic-wavelengths $\{\Lambda_D, \Lambda_S\}$ as,

$$\Lambda_D = \frac{\lambda_1 \lambda_2}{|\lambda_1 - \lambda_2|}; \quad for \quad \varphi_1 - \varphi_2 = \left(\frac{2\pi}{\lambda_1} - \frac{2\pi}{\lambda_2}\right)w = \frac{2\pi}{\Lambda_D}w;$$
$$\Lambda_S = \frac{\lambda_1 \lambda_2}{|\lambda_1 + \lambda_2|}; \quad for \quad \varphi_1 + \varphi_2 = \left(\frac{2\pi}{\lambda_1} + \frac{2\pi}{\lambda_2}\right)w = \frac{2\pi}{\Lambda_S}w. \quad (2)$$

The wavelength $\Lambda_D$ is longer than either $\{\lambda_1, \lambda_2\}$, or $\Lambda_D \gg \{\lambda_1, \lambda_2\}$. In contrast, $\Lambda_S$ is shorter than $\{\lambda_1, \lambda_2\}$; or $\Lambda_S \approx \lambda_1 / 2$ for $\lambda_2 = (1+\varepsilon)\lambda_1$; $\varepsilon \approx 0$.

Given that $\varphi_S = G\varphi_D$ ($G \gg 1.0$), the phase-sum $\varphi_S^W = W(\varphi_S)$ is highly wrapped ($W(x) = \text{Arg}[\exp(ix)]$). However $\varphi_S^W$ may be unwrapped from $\varphi_D$ by temporal phase-unwrapping, but it would need at least $G$ intermediate sensitive wrapped-phases [21-23]. To solve this Servin et al. [24,25] developed the 2-sensitivity phase-unwrapper,

$$\varphi_S = G\varphi_D + W\left[\varphi_S^W - G\varphi_D\right]; \quad \{\varphi_D \in (-\pi, \pi), (\varphi_S - G\varphi_D) \in (-\pi, \pi)\}. \quad (3)$$

This formula [24,25] predates the page-long algorithm in [31], and unwraps $\varphi_S^W$ directly from $\varphi_D \in (-\pi, \pi)$. This is the status in dual-wavelength phase-metrology as mid 2017.

## 3. Synthetic phase-difference and phase-sum estimations

Using the fringe-patterns in Eq. (1), and using the least-squares $M$-step PSA we demodulate our fringes obtaining the analytic signals [2],

$$A_1(x,y)e^{i\varphi_1(x,y)} = \sum_{m=0}^{M-1} I_1(x,y,m\Delta)e^{im\Delta},$$
$$A_2(x,y)e^{i\varphi_2(x,y)} = \sum_{m=0}^{M-1} I_2(x,y,m\Delta)e^{im\Delta}; \qquad \Delta = 2\pi/M. \tag{4}$$

Then the following products are computed,

$$A_1 e^{i\varphi_1}\left[A_2 e^{i\varphi_2}\right]^* = A_1 A_2 e^{i[\varphi_1-\varphi_2]}; \qquad A_1 e^{i\varphi_1}\left[A_2 e^{i\varphi_2}\right] = A_1 A_2 e^{i[\varphi_1+\varphi_2]}. \tag{5}$$

Obtaining,

$$\varphi_D = \frac{2\pi}{\Lambda_D}w\ ; \qquad \Lambda_D = \frac{\lambda_1\lambda_2}{|\lambda_1-\lambda_2|};$$
$$\varphi_S^W = W\left[\frac{2\pi}{\Lambda_S}w\right]; \qquad \Lambda_S = \frac{\lambda_1\lambda_2}{|\lambda_1+\lambda_2|}\ ; \tag{6}$$

Where $W(x) = \text{Arg}[\exp(ix)]$. Note that $\varphi_S^W = W[\varphi_1+\varphi_2]$ is highly wrapped and super-sensitive (beyond the optical range) while $\varphi_D \in (-\pi,\pi)$. Finally the sensitivity between $\{\varphi_S,\varphi_D\}$ is,

$$G = \frac{\varphi_S}{\varphi_D} = \frac{\lambda_1+\lambda_2}{|\lambda_1-\lambda_2|} = \frac{\Lambda_D}{\Lambda_S}, \tag{7}$$

That is, $\varphi_S(x,y)$ is $(\Lambda_D/\Lambda_S)$ times more sensitive than $\varphi_D(x,y)$. This result although implicit in [28,29,31], was not explicitly stated as we do here. The SNR and error-free unwrapping for $\varphi_S(x,y)$ are tightly tied to the sensitivity increase $G$.

## 4. SNR between the phase-sum and phase-difference

In practice the demodulated phases $\{\varphi_1,\varphi_2\}$ are corrupted by white noise $\{n_1,n_2\}$ [2] as,

$$\varphi_1(x,y) \to \varphi_1(x,y) + n_1(x,y);$$
$$\varphi_2(x,y) \to \varphi_2(x,y) + n_2(x,y). \tag{8}$$

Where $\{n_1,n_2\}$, ($n_2 \neq n_1$) are samples of a Gaussian random process [2,26]. The fields $\{n_1,n_2\}$ have zero average and variance $\sigma^2 = E\{n_1\} = E\{n_2\}$; being $E\{\cdot\}$ the ensemble average [26]. Then the phase-difference and phase-sum are,

$$\varphi_D = \varphi_2 - \varphi_1 = \left(\frac{2\pi}{\Lambda_D}\right)w + n_2 - n_1;$$
$$\varphi_S = \varphi_2 + \varphi_1 = \left(\frac{2\pi}{\Lambda_S}\right)w + n_2 + n_1. \tag{9}$$

In a Banach space, $(1/A_\Omega)\iint |f|^2 d\Omega$ represents the average-energy or power of $f$; being $A_\Omega$ the area of well-defined fringes $\Omega$. As a consequence the SNR for $\{\varphi_D,\varphi_S\}$ are,

$$\text{SNR}(\varphi_D) = \frac{Signal\,Power}{Noise\,Power} = \frac{\left(\frac{2\pi}{\Lambda_D}\right)^2 \iint\limits_{(x,y)\in\Omega} |w|^2 \, d\Omega}{\iint\limits_{(x,y)\in\Omega} |n_2 - n_1|^2 \, d\Omega};$$

$$\text{SNR}(\varphi_S) = \frac{Signal\,Power}{Noise\,Power} = \frac{\left(\frac{2\pi}{\Lambda_S}\right)^2 \iint\limits_{(x,y)\in\Omega} |w|^2 \, d\Omega}{\iint\limits_{(x,y)\in\Omega} |n_2 + n_1|^2 \, d\Omega}.$$

(10)

Assuming $\{n_1, n_2\}$ ergodic and stationary, the noise-power of $\{n_2 - n_1, n_2 + n_1\}$ are equal [26] so the SNR gain between $\varphi_S(x,y)$ and $\varphi_D(x,y)$ is,

$$\frac{\text{SNR}(\varphi_1 + \varphi_2)}{\text{SNR}(\varphi_1 - \varphi_2)} = \left(\frac{\Lambda_D}{\Lambda_S}\right)^2 = G^2. \tag{11}$$

Thus, the super-sensitive phase $\varphi_S(x,y)$ has $G^2$ higher SNR than $\varphi_D(x,y)$. So it is much better use $\varphi_S(x,y)$ instead of just $\varphi_D(x,y)$. As far as we know, this result has not been published before, in particular [28,29,31] do not contain any SNR mathematical analysis.

## 5. Error-free phase-unwrapping for discontinuous phase-sum

Here we unwrap the noisy $\varphi_S^W = W[\varphi_1 + \varphi_2 + n_S]$ from noisy $\varphi_D = \varphi_1 - \varphi_2 + n_D$ without errors. The high gain $G \gg 1.0$ translates into noise limitations for error-free unwrapping for $\varphi_S^W$. Here we are assuming small phase-noise i.e. $\{|n_S| \ll \pi, |n_D| \ll \pi\}$.

For the noisy $(\varphi_D + n_D)$ and $(\varphi_S^W + n_S)$ our temporal unwrapping formula is [24, 25],

$$\varphi_S + n_S = G(\varphi_D + n_D) + W\left[(\varphi_S^W + n_S) - G(\varphi_D + n_D)\right]; \qquad W[x] = \text{Arg}(e^{ix}). \tag{12}$$

The conditions for this phase-unwrapper to work properly are (reported in [24,25]),

$$(\varphi_D + n_D) \in (-\pi, \pi), \quad \text{and} \quad \left[\varphi_S + n_S - G(\varphi_D + n_D)\right] \in (-\pi, \pi). \tag{13}$$

Substituting the noiseless sensitivity relation $\varphi_S - G\varphi_D = 0$ into the second condition,

$$(n_S - Gn_D) \in (-\pi, \pi); \qquad G = \frac{\Lambda_D}{\Lambda_S}. \tag{14}$$

This noisy signal must reside within $(-\pi, \pi)$ to unwrap $\varphi_S^W + n_S$ without errors. This last condition is a new result not reported in [24,25], neither in [28-32].

The most important fact to take note here is the product $Gn_D$ ($G \gg 1.0$). Therefore we must keep $n_D$ as low as possible by using many-steps PSAs. And as we said, this useful result has not been reported before [1-32].

## 6. Application example to super-resolution fringe-projection profilometry

Here we exemplify the general theory for dual-wavelength super-sensitive phase-metrology using two examples from 3D profilometry. These profilometry examples fully agree with the three previous experiments that use the phase-sum in other phase-metrology areas [28,29,31].

Digital fringe-projection profilometry has been known for many years [1]. A set up for three-dimensional (3D) profilometry of solids is shown in Fig. 1.

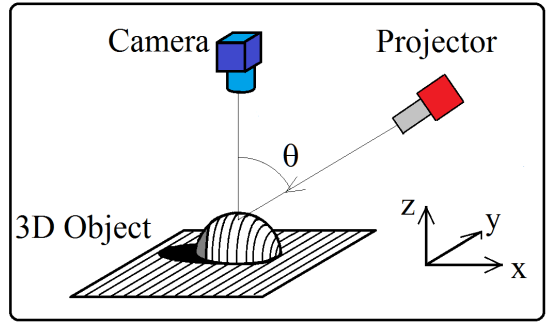

Fig. 1. Schematic of a fringe-projection profilometer. The digitizing solid is a sphere segment and $\theta$ is the sensitivity angle of this profilometer.

The mathematical model for the fringes $I_1(x,y)$ and $I_2(x,y)$ taken by the CCD camera are,

$$I_1(x,y,m) = a(x,y) + b(x,y)\cos\left[u_1\tan(\theta)h(x,y) + u_1 x + \frac{2\pi}{4}m\right];$$
$$I_2(x,y,m) = a(x,y) + b(x,y)\cos\left[u_2\tan(\theta)h(x,y) + u_2 x + \frac{2\pi}{4}m\right]; \quad m = \{0.1.2.3\}. \quad (15)$$

Where $h(x,y)$ the solid's height and $\theta$ the sensitivity angle (Fig. 1); the carriers $\{u_1,u_2\}$ are in radians/pixel. The angle $(\theta)$ is fixed and small to avoid large self-occluding shadows. The upper limit for $\{u_1,u_2\}$ is the Nyquist rate of $\pi$ radians/pixel. However, here we are not using very-high carriers because the fringes would be hardly observable. The sensitivity gain is,

$$G = \frac{\varphi_S}{\varphi_D} = \frac{u_1 + u_2}{|u_1 - u_2|}. \quad (16)$$

All previous results apply just by substituting $\Lambda_D/\Lambda_S \to (u_1+u_2)/|u_1-u_2|$.

*6.1 A calibrating spherical metallic cap*

The spherical metallic cap in Fig. 2 is used as calibrating solid. This is a regular photography, not a three-dimensional (3D) digital rendering.

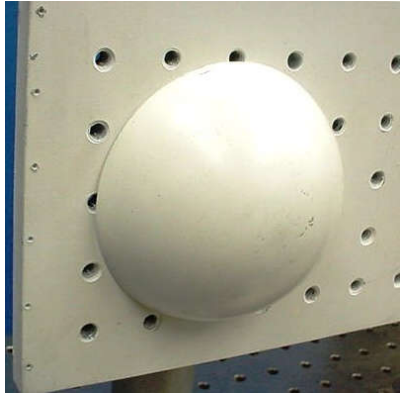

Fig. 2. Regular digital photograph of a calibrating spherical metallic cap..

In Fig. 3 we show the two close spatial-frequencies (close phase-sensitivities) linear-fringes projected on this spherical object. In the upper row of Fig. 3, we have projected linear-fringes with spatial-frequency $u_1 = 2\pi/7$ radians/pixel, while in the lower row the fringes have a carrier of $u_2 = 2\pi/6$ radians/pixel.

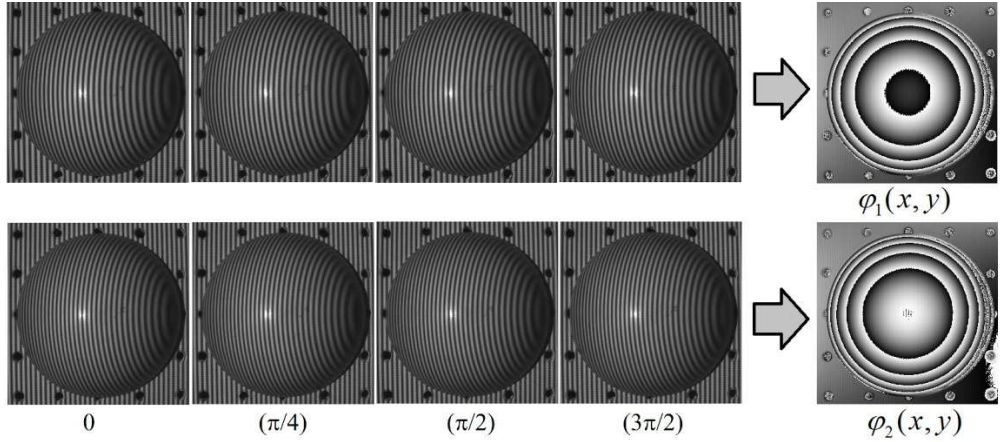

0      (π/4)      (π/2)      (3π/2)      $\varphi_2(x,y)$

Fig. 3. A spherical segment illuminated by linear fringes and their phase demodulation. The phase-shifting among the fringe-patterns are {0,π/4,π/2,3π/2}. The upper row has lower phase-sensitivity fringes than the lower row. The right column shows the demodulated wrapped phases.

Figure 4 shows the phase-subtraction (upper row) and the phase-addition (lower row) of the two demodulated phases in Fig. 4. The sensitivity gain $G$ is then given by,

$$G = \frac{u_1 + u_2}{|u_1 - u_2|} = \frac{(2\pi/6 + 2\pi/7)\,Radians/Pixel}{|2\pi/6 - 2\pi/7|\,Radians/Pixel} = 13, \qquad (17)$$

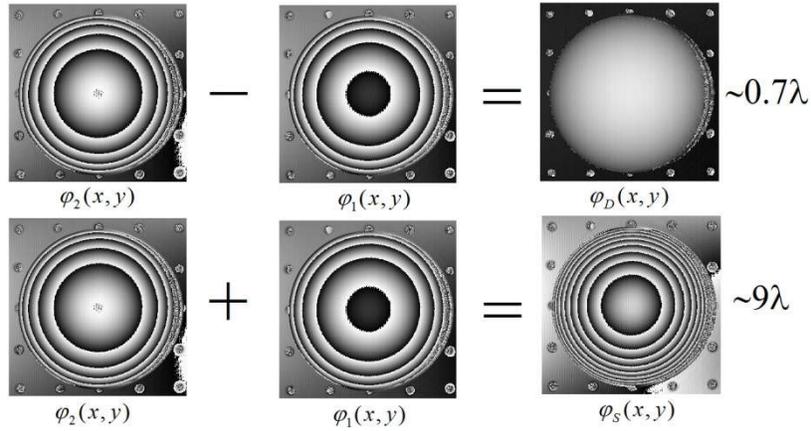

Fig. 4. The upper row shows the phase-difference while the lower row the phase-sum. The phase difference has about 0.7 lambda sensitivity and non-wrapped. The phase-sum is however wrapped about 9-times.

That is, we have 13-times more sensitivity in the phase-sum $\varphi_S(x,y)$ compared to the non-wrapped phase-difference $\varphi_D(x,y)$. Figure 5 shows the unwrapped phase-sum $\varphi_S(x,y)$ taking as first estimation the non-wrapped $\varphi_D(x,y)$.

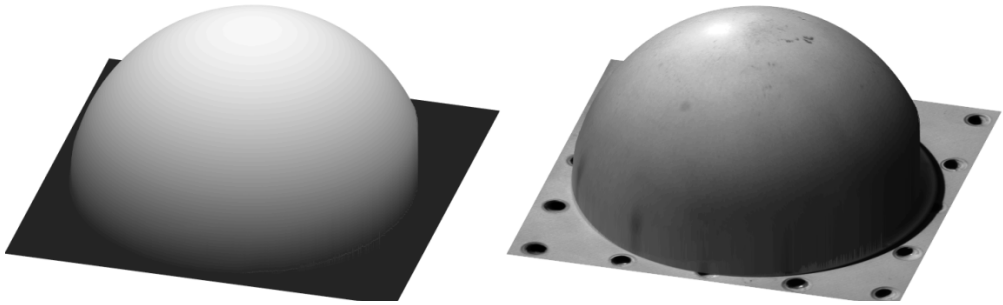

Fig.5. Three-dimensional (3D) digital rendering of the recovered unwrapped phase-sum of the spheric solid. The left-side shows the gray-coded phase-sum, while on the right-side shows the phase-sum with the solid's photograph as texture.

In Fig. 6 we show a central phase-cut of $\varphi_D(x,y)$ and $\varphi_S(x,y)$ to gauge the different amounts of degrading phase-noise.

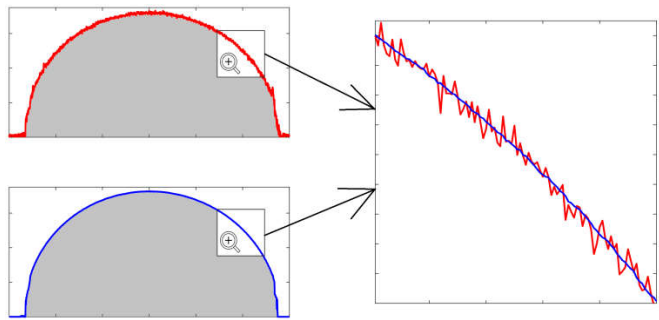

Fig. 6. A central-cut of the digitized spherical cap. The red trace is the phase-difference and the blue trace is the unwrapped phase-sum. The blue trace has significant lower phase-noise.

In the zoomed-in detail one may see the difference in phase-noise amplitude corresponding to the noisier phase-difference in red than the blue phase-sum.

*6.2 A fluorescent spiral light bulb*

Figure 7 shows our next object a spiral fluorescent lamp. This is a regular photography not a digital rendering.

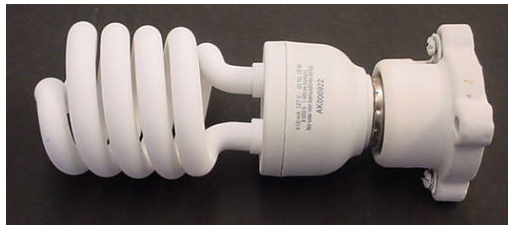

Fig. 7. Regular Photograph (not 3D rendering) of a spiral fluorescent lamp under analysis.

Figure 8 shows phase-shifted linear-fringes with low and high carriers. The upper row carrier is $u_1 = 2\pi/9$ radians/pixel, while the lower row carrier is $u_2 = 2\pi/8$ radians/pixel.

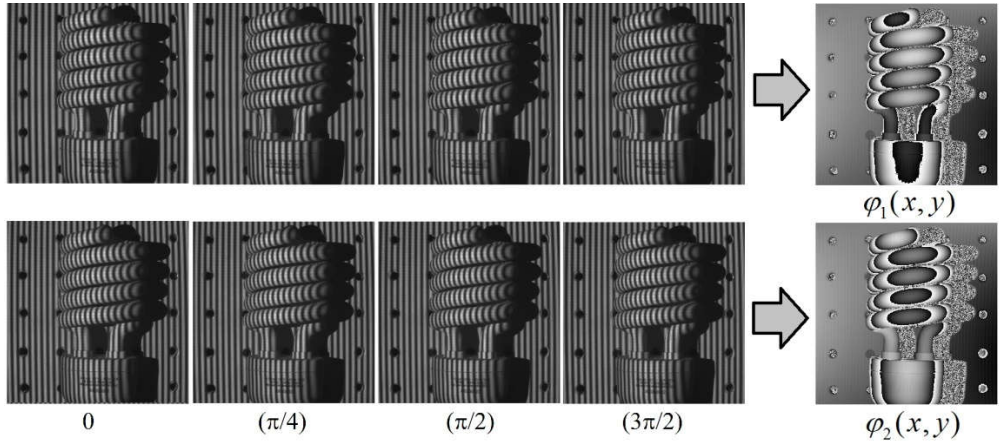

| 0 | (π/4) | (π/2) | (3π/2) |

Fig. 8. The spiral lamp illuminated by two close carrier-frequency fringes. The lower carrier fringes are shown in the first row, while the higher carrier fringes are shown in the second row. Their 4-step PSA demodulated wrapped phase are shown at the right column.

Figure 9 shows the phase-difference $\varphi_D(x,y)$ and phase-sum $\varphi_S(x,y)$ of the two sensitivities demodulated phase-maps in Fig. 9.

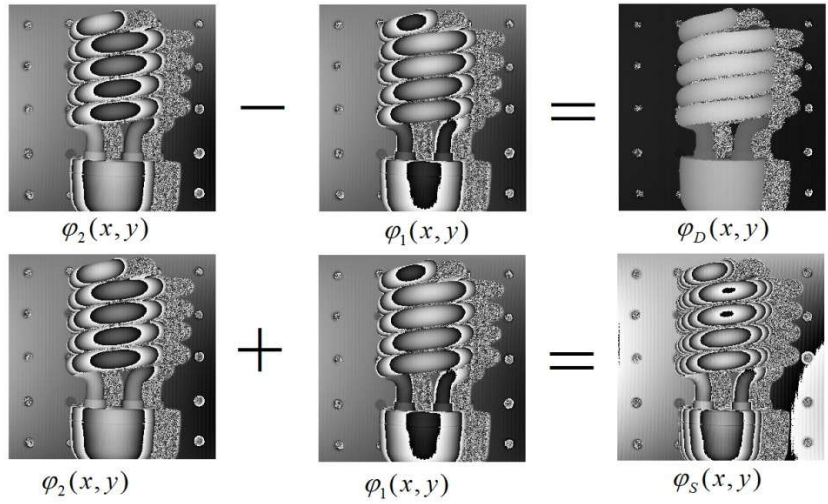

Fig. 9. The upper row shows the non-wrapped phase-difference while the lower row shows the highly wrapped phase-sum. The phase-sum has much higher sensitivity, so it is highly wrapped.

The sensitivity gain $G$ in the spiral-lamp case is given by,

$$G = \frac{u_1+u_2}{|u_1-u_2|} = \frac{(2\pi/8+2\pi/9)\,Radians/Pixel}{|2\pi/8-2\pi/9|\,Radians/Pixel} = 17, \qquad (18)$$

That is, we have 17-times more sensitivity in the phase-sum $\varphi_S(x,y)$ compared to the phase-difference $\varphi_D(x,y)$.

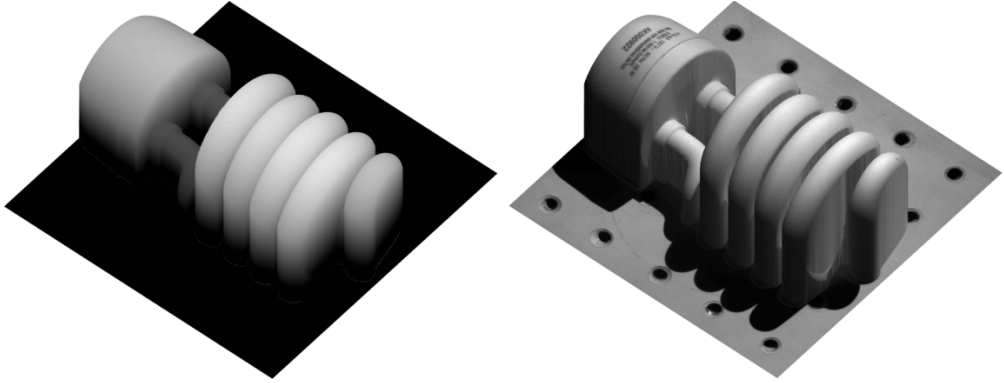

Fig. 10. Digital rendering of the recovered unwrapped super-sensitive phase-sum. The left-side rendering is gray-coded phase. For the right-side we used its photograph as rendering texture.

Figure 10 shows the unwrapped super-sensitive phase-sum.

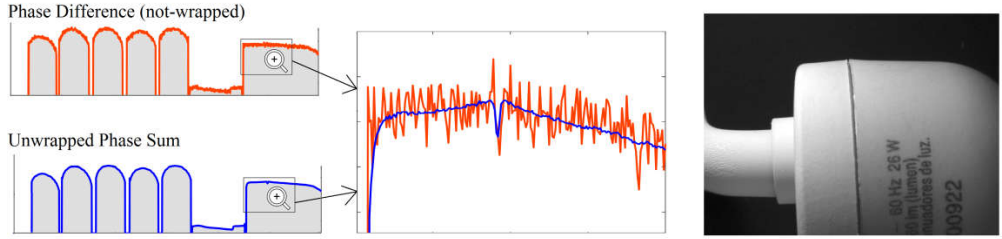

Fig. 11. Comparison of the SNR between the phase-difference (in red), and the super-sensitive phase-sum (in blue). We have zoomed-in to clearly see the depression at the joining of the two plastic pieces shown in the photograph-detail.

Figure 11 shows two phase-cuts of the digitized spiral fluorescent lamp. In the far right a regular zoom-in photograph is shown to see the joint between these two plastic pieces. This joint depression is hardly seen in the red-graph because it is immersed in measuring noise.

## 7. SNR for high-carrier frequency single and phase-sum profilometry

Before summarizing, we comment another widely used dual-wavelength profilometry strategy [24,25]. This consist of using a very-low carrier $u_{Low}$ (i.e. 512 pixels/period) to obtain absolute phase, plus a very-high one $u_{High} = \pi - |\varepsilon|$ ($\varepsilon \approx 0$) to detect fine surface details. The absolute-phase $\varphi_{Low} = u_{Low} \tan(\theta) h$ is used to unwrap $\varphi_{High} = u_{High} \tan(\theta) h$ [24,25]. The unwrapped phases for single and phase-sum high-carrier linear fringes are,

$$\varphi_{High} = u_{High} \tan(\theta) h;$$
$$\varphi_1 + \varphi_2 = (u_1 + u_2) \tan(\theta) h. \qquad (19)$$

Assuming that $u_{High} = (u_1 + u_2)/2$, thus $u_1 + u_2 = 2u_{High}$, and the phase-sum $\varphi_S$ has twice the sensitivity of $\varphi_{High}$,

$$\frac{\varphi_S}{\varphi_{High}} = \frac{u_1 + u_2}{u_{High}} = 2. \qquad (20)$$

And the SNR gain between $(\varphi_1 + \varphi_2)$ and $\varphi_{High}$ is,

$$\frac{\text{SNR}(\varphi_1+\varphi_2)}{\text{SNR}(\varphi_{High})} = \left(\frac{u_1+u_2}{u_{High}}\right)^2 = (2)^2 = 4. \tag{21}$$

That is, the SNR for $\varphi_S = \varphi_1 + \varphi_2$ is four times higher than projecting a single fringe-pattern with carrier $u_{High} = (u_1+u_2)/2$; $u_{High} = \pi - |\varepsilon|$. Therefore the synthetic phase-sum doubles the sensitivity and quadruple the SNR with respect to projecting a single high-frequency fringe-pattern. Therefore there is a clear advantage of using two high-carrier fringes for the same total number of fringe patters; assuming four phase-shifted fringes to obtain $\varphi_{Low} = u_{Low} \tan(\theta) h$ plus another four to obtain $\varphi_{High} = u_{High} \tan(\theta) h$. This brief analysis was included following a useful suggestion of a reviewer..

## 8. Summary

We have presented a general mathematical theory for super-sensitivity two-wavelength phase-metrology. In particular we have presented close-form formulae to quantify the SNR and error-free phase-unwrapping for the super-sensitive phase-sum. In particular we have applied this general two-wavelength metrology theory to fringe-projection profilometry. This profilometry method uses linear fringe-patterns with close-sensitivity $\varphi_1 \approx \varphi_2$. Then we compute $\varphi_D = \varphi_2 - \varphi_1$ and $\varphi_S = \varphi_1 + \varphi_2$; being the sensitivity between $\{\varphi_1, \varphi_1\}$ close enough so their difference is non-wrapped i.e. $(\varphi_2 - \varphi_1) \in (-\pi, \pi)$. In contrast the super-sensitivity $\varphi_S = \varphi_1 + \varphi_2$ is highly wrapped. Finally we adapted our previous [24,25] phase-unwrapper to unwrap $\varphi_S(x,y)$ error-free directly from $\varphi_D(x,y)$, using a single formula (Eqs. (12)).

As far as we know, the only previous studies [28,29,31] that use $\varphi_S = \varphi_1 + \varphi_2$ and $\varphi_D = \varphi_2 - \varphi_1$ do not contain closed-form mathematical formulae for the SNR of $\{\varphi_S, \varphi_D\}$, neither a mathematical formula for *error-free* unwrapping for $\varphi_S = \varphi_1 + \varphi_2$. Here in contrast, we present closed-form formulae for the SNR of $\{\varphi_1+\varphi_2, \varphi_1-\varphi_2\}$, and to unwrap $\varphi_S^W$ without errors. At the risk of being repetitive, in [28-31] no mathematical analysis and closed-form formulas are provided for: the SNR of $\{\varphi_D, \varphi_S\}$, and for unwrapping $\varphi_S^W$ without errors. Just specific numerical/experimental examples were displayed [28-31]. However, we know that experimental/numerical demonstrations and/or plausible conjectures, as numerous as they might be, never substitute rigorously obtained mathematical formulae.

**Acknowledgments**

The authors acknowledge Cornell University for supporting the e-print open repository arXiv.org, and the permission granted by the Optical Society of America which allows uploading submitted manuscripts to arXiv.org. We thank a reviewer for calling our attention to the recent work by Di et al. work [29]. Using Google Scholar we tracked down its two related works [28,31].